\documentclass[aps,prl,twocolumn,groupedaddress]{revtex4}

\newcommand{\bfc}{{\bf c}}
\newcommand{\bfci}{\bfc_i}
\newcommand{\bfu}{{\bf u}}
\newcommand{\bfx}{{\bf x}}
\newcommand{\Neq}[1]{{N^{\mbox{\scriptsize eq}}_{{#1}}}}
\newcommand{\dt}{{\Delta t}}
\newcommand{\bfbeta}{{\mbox{\boldmath $\beta$}}}
\newcommand{\bfone}{{\mbox{\boldmath $1$}}}
\newcommand{\bfnabla}{{\mbox{\boldmath $\nabla$}}}
\newcommand{\taumin}{{\tau_{\mbox{\tiny $\min$}}}}
\newcommand{\rhoomb}{{\frac{\rho}{mb}}}
\newcommand{\phimu}{{\phi\left(x\right)}}
\newcommand{\phipmu}{{\phi'\left(x\right)}}
\newcommand{\phippmu}{{\phi''\left(x\right)}}
\newcommand{\phipx}{{\phi'\left(x\right)}}
\newcommand{\phippx}{{\phi''\left(x\right)}}

\begin{document}


\title{Galilean-Invariant Lattice-Boltzmann Models with H-Theorem}


\author{Bruce M. Boghosian, Peter J. Love}
\affiliation{Department of Mathematics, Bromfield-Pearson Hall, Tufts University, Medford, MA 02155, U.S.A.}
\author{Peter V. Coveney}
\affiliation{Centre for Computational Science, Department of Chemistry, University College London, London, 20 Gordon St., WC1H 0AJ, UK}

\author{Iliya V. Karlin}
\affiliation{ETH Z\"{u}rich, Department of Materials, Institute of Polymers, ETH-Zentrum, Sonneggstr. 3, ML J 19, CH-8092 Z\"{u}rich, Switzerland}

\author{Sauro Succi}
\affiliation{Istituto Applicazioni Del Calcolo, viale del Policlinico 137, 00161 Roma, Italy}
\author{Jeffrey Yepez}
\affiliation{Air Force Research Laboratory, Hanscom A. F. B., MA 01731-3010,U. S. A.}


\date{\today}

\begin{abstract}
We demonstrate that the requirement of galilean invariance determines
the choice of $H$ function for a wide class of entropic lattice
Boltzmann models for the incompressible Navier-Stokes equations.  The
required $H$ function has the form of the Burg entropy for $D=2$, and
of a Tsallis entropy with $q=1-\frac{2}{D}$ for $D>2$, where $D$ is
the number of spatial dimensions.  We use this observation to
construct a fully explicit, unconditionally stable, galilean
invariant, lattice-Boltzmann model for the incompressible
Navier-Stokes equations, for which attainable Reynolds number is
limited only by grid resolution.
\end{abstract}


\maketitle

\section{Introduction}

Lattice Boltzmann models of fluids~\cite{bib:bsv,bib:succi} evolve a
single-particle distribution function in discrete time steps on a
regular spatial lattice, with a discrete velocity space comprised of
the lattice vectors themselves.  The single-particle distribution
corresponding to lattice vector $\bfci$ at lattice position $\bfx$ and
time step $t$ is denoted by $N_i(\bfx,t)$.  The simplest variety of
lattice Boltzmann models employ a BGK operator~\cite{bib:qian}, so
that their evolution equation is
\[
N_i(\bfx+\bfci,t+\dt) = N_i(\bfx,t) + \frac{1}{\tau}\left[\Neq{i}(\bfx,t) - N_i(\bfx,t)\right]
\]
for $i=1,\ldots,b$.  Here $b$ is the coordination number of the
lattice, $\Neq{i}(\bfx,t)$ is a specified equilibrium distribution
function that depends only on the values of the conserved quantities
at a site, and $\tau$ is a characteristic collisional relaxation time.
Using the Chapman-Enskog analysis it is possible to show that the mass
and momentum moments of the distribution function will obey the
Navier-Stokes equations for certain choices of equilibrium
distribution~\cite{bib:succi}.

The viscosity that appears in the Navier-Stokes equations obtained
from these models is proportional to $\tau-\mbox{\tiny
$\frac{1}{2}$}$.  To lower viscosity and thereby increase Reynolds
number, practitioners often over-relax the collision operator by using
values of $\tau$ in the range $(\mbox{\tiny $\frac{1}{2}$}, 1]$.  For
sufficiently small $\tau$, however, the method loses numerical
stability, and this consideration limits the lowest Reynolds numbers
attainable.

In an effort to understand these instabilities, there has been much
recent interest in {\it entropic lattice Boltzmann
models}~\cite{bib:karlina,bib:karlinb,bib:entropicLB}.  These models
are motivated by the fact that the loss of stability is due to the
absence of an $H$ theorem.  Numerical instabilities evolve in ways
that would be precluded by the existence of a Lyapunov function.  The
idea behind entropic lattice Boltzmann models is to specify an $H$
function, rather than just the form of the equilibrium.  Of course,
the equilibrium distribution will be that which extremizes the $H$
function.  The evolution will be required never to decrease $H$,
yielding a rigorous discrete-time $H$-theorem; this is to be
distinguished from other discrete models of fluid dynamics for which
an $H$-theorem may be demonstrated only in the limit of vanishing time
step~\cite{bib:dpd7}.

To ensure that collisions never decrease $H$, the collision time
$\tau$ is made a function of the incoming state by solving for the
smallest value $\taumin<1$ that does not increase $H$.  The value then
used is $\tau = \taumin/\kappa$ where $0<\kappa<1$.  It has been shown
that the expression for the viscosity obtained by the Chapman-Enskog
analysis will approach zero as $\kappa$ approaches
unity~\cite{bib:karlina,bib:karlinb,bib:entropicLB}.  Thus, the
entropic lattice Boltzmann methodology allows for arbitrarily low
viscosity together with a rigorous discrete-time $H$ theorem, and thus
absolute stability.  The upper limit to the Reynolds numbers
attainable by the model is therefore determined by loss of resolution
of the smallest eddies, rather than by loss of
stability~\cite{bib:karlinc,bib:karlind}.

In a recent review of the subject, Succi, Karlin and
Chen~\cite{bib:skc} have pointed out that entropic lattice Boltzmann
models have three important desiderata: Galilean invariance,
non-negativity of the distribution function, and ease of determining
the local equilibrium distribution at each site at each timestep.

In this paper, we shall construct entropic lattice Boltzmann models
for the incompressible Navier-Stokes equations which are gallilean
invariant to second order in the Mach number expansion of the
distribution function (quasi-perfect in the terminology
of~\cite{bib:skc}).  We shall show that the requirement of galilean
invariance makes the choice of $H$ function unique.  We shall show that the required function has the
form of the Burg entropy~\cite{bib:karline} in two dimensions, and the
Tsallis entropy in higher dimensions.  While the analogous problem for
the compressible Navier-Stokes equations is difficult and remains
outstanding, the purpose of this paper is to point out that the
incompressible case is nontrivial and interesting in its own right.

\section{Equilibrium Distribution}

We consider a Bravais lattice of coordination number $b$ in $D$
dimensions.  We denote the lattice vectors by $\bfci$ where
$i=1,\ldots,b$, and their magnitudes by $c=\left|\bfci\right|$.  We
demand that the lattice symmetry group be sufficiently large that the
only fourth-rank tensors that are invariant under its group action are
isotropic.  The mass and momentum densities are given by
\begin{equation}
\rho = \sum_{i=1}^b m N_i
\label{eq:rho}
\end{equation}
and
\begin{equation}
\rho\bfu = \sum_{i=1}^b m\bfci N_i,
\label{eq:rhou}
\end{equation}
where $m$ is the particle mass, and $\bfu$ is the hydrodynamic
velocity $D$-vector.  These $D+1$ quantities must be conserved in
collisions.

If we regard the $N_i$, for $i=1,\ldots,b$, as coordinates in a
$b$-dimensional space, the conservation laws (\ref{eq:rho}) and
(\ref{eq:rhou}) restrict the collision outcomes to a $b-(D+1)$
dimensional subspace.  Since the conserved quantities are linear
functions of the $N_i$'s, the nonnegativity requirement
\begin{equation}
N_i\geq 0
\label{eq:constraint}
\end{equation}
is satisfied within a compact polytope whose faces are given by the
$b$ equations $N_i=0$ for $i=1,\ldots,b$.  We assume that the
$H$-function is of trace form
\[
H
= \sum_{i=1}^b h\left(N_i\right),
\]
where $h'(x)\geq 0$ for $x>0$.  If $\lim_{x\rightarrow
0}h'(x)=\infty$, then the normal derivative of $H$ goes to
negative infinity on the polytope boundary, enforcing the
nonnegativity constraint, Eq.~(\ref{eq:constraint}).  The purpose of
this paper is to demonstrate that the requirement of galilean
invariance uniquely determines the choice of function $h(x)$.

The equilibrium distribution function may be found by extremizing $H$
with respect to the $N_i$, subject to the constraints,
Eqs.~(\ref{eq:rho}) and (\ref{eq:rhou}),
\[
0 = \frac{\partial}{\partial N_i}\left(H-\frac{\mu}{m}\rho-\frac{\bfbeta}{m}\cdot\rho\bfu\right),
\]
where $\mu/m$ and $\bfbeta/m$ are Lagrange multipliers.  We quickly
find
\[
0 = h'\left(N_i\right) - \mu - \bfbeta\cdot\bfc_i,
\]
and so
\begin{equation}
\Neq{i} = \phi\left(\mu + \bfbeta\cdot\bfc_i\right),
\label{eq:exacteq}
\end{equation}
where the function $\phi$ is the inverse function of $h'$.  The
constants $\mu$ and $\bfbeta$ are determined by Eqs.~(\ref{eq:rho})
and (\ref{eq:rhou}), though it is generally impossible to find an
exact analytic expression for them in terms of the conserved
quantities $\rho$ and $\rho\bfu$; rather one must solve for them
numerically or perform a Taylor expansion in Mach number.  We adopt
the latter approach below.

\section{Galilean Invariance}

We seek to Taylor expansion the equilibrium distribution in Mach
number because (i) we can do so analytically, (ii) only the first two
terms of that expansion determine the form of the incompressible
Navier-Stokes equations, and (iii) that expansion is a useful initial
guess for any numerical solution.  From general symmetry arguments it
is clear that $\bfbeta$ will be proportional to the hydrodynamic
velocity $\bfu$, so that we may begin our Mach number expansion by
expanding Eq.~(\ref{eq:exacteq}) for small $\bfbeta$.  We get
\[
\Neq{i} = \phi(\mu) + \phi'(\mu)\bfbeta\cdot\bfc_i + \frac{1}{2}\phi''(\mu)\bfbeta\bfbeta : \bfc_i\bfc_i +\cdots
\]
Inserting this into Eqs.~(\ref{eq:rho}) and (\ref{eq:rhou}), and using
general properties of the Bravais lattice, we find
\[
\rho = mb\phi(\mu) + \frac{mbc^2}{2D}\phi''(\mu)\beta^2 + \cdots
\]
and
\[
\rho\bfu = \frac{mbc^2}{D}\phi'(\mu)\bfbeta + \cdots,
\]
where the ellipses denote third or higher order in Mach number.
Inverting this perturbatively we find that, to second order in Mach
number, the Lagrange multipliers are given by
\[
\mu = x -
\frac{D}{2c^2}
\frac{\left(\rhoomb\right)^2 \phippmu}{\left[\phipmu\right]^2}  u^2 + \cdots,
\]
where $x\equiv h'\left(\rhoomb\right)$, and by
\[
\bfbeta =
\frac{D}{c^2}\frac{\rhoomb}{\phipmu}\bfu + \cdots.
\]
Inserting these into Eq.~(\ref{eq:exacteq}), we obtain the equilibrium
distribution,
\begin{widetext}
\begin{equation}
\Neq{i} =
\rhoomb\left[
1 + \frac{D}{c^2}\bfci\cdot\bfu + \frac{D^2}{2c^4}
\frac{\phimu\phippmu}{\left[\phipmu\right]^2}
\left(\bfci\bfci - \frac{c^2}{D}\bfone\right) : \bfu\bfu + \cdots
\right]
\label{eq:nneqa}
\end{equation}
Now it is well known that a Chapman-Enskog analysis based on the
equilibrium distribution
\begin{equation}
\Neq{i} =
\rhoomb\left[
1 + \frac{D}{c^2}\bfci\cdot\bfu + \frac{D(D+2)}{2c^4}
g
\left(\bfci\bfci - \frac{c^2}{D}\bfone\right) : \bfu\bfu +
\cdots
\right]
\label{eq:nneqb}
\end{equation}
\end{widetext}
will give rise to the incompressible Navier-Stokes equations
\[
\bfnabla\cdot\bfu = 0
\]
and
\[
\frac{\partial\bfu}{\partial t} + g \bfu\cdot\bfnabla\bfu =
-\frac{1}{\rho}\bfnabla P + \nu\nabla^2\bfu.
\]
Comparing Eqs.~(\ref{eq:nneqa}) and (\ref{eq:nneqb}), we identify
\[
g = \left(\frac{D}{D+2}\right)
\frac{\phimu\;\phippmu}
{\left[\phipmu\right]^2}.
\]
The factor $g$ destroys the form of the convective derivative, and
hence breaks galilean invariance.  To recover galilean invariance we
demand that $g=1$, and this yields the second-order nonlinear
differential equation
\[
\phi(x)\phippx = \left(1+\frac{2}{D}\right)\left[\phipx\right]^2.
\]
The general solution to this equation is of the form
\[
\phi(x) = C^{D/2} (x-aC)^P,
\]
where $C$ and $a$ are arbitrary constants, and $P$ is to be
determined.  We quickly find that $P$ must be either $0$ or $-D/2$.
Since a constant $\phi$ would not yield a well defined $h'$, we see
that we must have $\phi(x)=C^{D/2} (x-aC)^{-D/2}$, whence $h'(x) =
C(a+ x^{-2/D})$, and this integrates to give
\begin{equation}
h(x) =
\left\{
\begin{array}{ll}
h_0 + C\left[ax + \ln x\right] &
\mbox{if $D=2$} \\
h_0 + C\left[ax+\left(\frac{x^{1-2/D}-1}{1-2/D}\right)\right] &
\mbox{if $D\neq 2$,}
\end{array}
\right.
\label{eq:zeta}
\end{equation}
where $h_0$ is constant.  In fact, the only effect of nonzero $h_0$ is to introduce an additive constant to $H$, and the only effect of nonunity $C$ is to scale $H$ by a constant factor.  In other words, $h(x)$ is uniquely specified only to within additive and multiplicative constants.  With this understanding, we may say that the requirement of galilean invariance has uniquely specified the choice of $H$.

In passing, we note that $\lim_{x\rightarrow 0} h'(x)=\infty$.  Thus
the nonnegativity constraint will be enforced by the dynamics.

Finally, we write the global Lyapunov function ${\cal H}\equiv\sum_{\bf x} H$ by summing
$h(N_i(\bfx,t))$ over the lattice.  Since the total mass is
conserved we have complete freedom to choose $a$, and so to within
additive and multiplicative constants ${\cal H}$ may be written
\[
{\cal H}(t)
\propto
\left\{
\begin{array}{ll}
\sum_\bfx\sum_i \ln\left[N_i(\bfx,t)\right] & \mbox{for $D=2$}\\
\sum_\bfx\sum_i \frac{\left[N_i(\bfx,t)\right]^{1-2/D}-N_i(\bfx,t)}{2/D} & \mbox{for $D\neq 2$,}
\end{array}
\right.
\]
for appropriate choices of $a$ and $C$.  This has the form of a Burg entropy~\cite{bib:karline} for $D=2$, and
a Tsallis entropy~\cite{bib:tsallis} with parameter
\[
q=1-\frac{2}{D}
\]
for $D\neq 2$.  We note that $D\leq 2$ corresponds to $q\leq 0$, and
$D>2$ corresponds to $q>0$.  It is interesting that it is only in the
infinite-dimensional limit, $D\rightarrow\infty$, where the set of
velocities becomes infinite, that $q\rightarrow 1$ and we recover the
Boltzmann-Gibbs entropy~\footnote{This limit must be taken
with great caution, since the sound speed, $c_s = c/\sqrt{D}$, also
vanishes in this limit, threatening the validity of the expansion in
Mach number $M=u/c_s$; we could, for example, take $u/c\sim D^{-3/4}$
so that $M\rightarrow 0$ even as $D\rightarrow\infty$.}.

The appearance of the Burg and Tsallis entropies in this context is
fascinating.  In a footnote of their recent review, Succi, Karlin and
Chen~\cite{bib:skc} noted that the entropy that gave rise to the
above-mentioned solvable model for a compressible fluid was related to
the Tsallis entropy with $q=3/2$, so there may be more than one
connection with Tsallis thermostatistics~\cite{bib:tsallis} lurking
here.  There are precious few situations in which the origins of
Tsallis thermostatistics can be traced analytically to an underlying
microscopic model, as we have done here.

\section{Conclusions}

We have presented galilean-invariant, entropic lattice Boltzmann
models for the incompressible Navier-Stokes equations.  We expect that
these models will be useful for the simulation of two- and
three-dimensional turbulence.  As noted by Succi, Karlin and
Chen~\cite{bib:skc}, the problem of finding perfect models for lattice
models of the {\it compressible} Navier-Stokes equations is much more
difficult and may well be impossible.  We found it interesting that
the simpler problem, for incompressible fluids, is itself very
nontrivial and interesting.  In particular, the appearance of the Burg
and Tsallis entropies for the $H$ function is surprising.  These
entropies have heretofore been associated with long-range
interactions, long-time memory or a fractal space-time structure.
This work indicates that they may also be relevant to models with
discretized space-time, and this surely warrants future study.

\section*{Acknowlegements}

BMB was supported in part by the U.S. Air Force Office of Scientific
Research under grant number F49620-01-1-0385, and in part by MesoSoft
Corporation.  He performed a portion of this work while at the Centre
for Computational Science, Department of Chemistry, Queen Mary,
University of London as an EPSRC Visiting Fellow under RealityGrid
contract GR/R67699.  Peter Love was supported by the DARPA QuIST
program under AFOSR grant number F49620-01-1-0566.


\end{document}